\documentclass[useAMS,usenatbib]{mn2e}
\usepackage{graphicx,rotate,url,mathptmx,lscape,times, color,subfigure}
\def\gtsim{\mathrel{\hbox{\rlap{\hbox{\lower4pt\hbox{$\sim$}}}\hbox{$>$}}}}
\def\lesssim{\mathrel{\hbox{\rlap{\hbox{\lower4pt\hbox{$\sim$}}}\hbox{$<$}}}}

\def\Msun{M$_{\odot}$}
\def\km{{\rm\thinspace km}}

\def\Mpc{{\rm\thinspace Mpc}}
\def\Msun{\hbox{$\rm\thinspace M_{\odot}$}}

\def\ps{{\rm\thinspace s^{-1}}}

\newcommand{\lsim}{\raisebox{-0.3ex}{\mbox{$\stackrel{<}{_\sim} \,$}}}
\newcommand{\gsim}{\raisebox{-0.3ex}{\mbox{$\stackrel{>}{_\sim} \,$}}}

\def\simless{\mathbin{\lower 3pt\hbox
	{$\,\rlap{\raise 5pt\hbox{$\char'074$}}\mathchar"7218\,$}}} 
\def\simgreat{\mathbin{\lower 3pt\hbox
	{$\,\rlap{\raise 5pt\hbox{$\char'076$}}\mathchar"7218\,$}}} 

\def\kmps{\hbox{$\km\ps\,$}}
\def\kmpspMpc{\hbox{$\kmps\Mpc^{-1}\,$}}

\def\h0{\hbox{{\rm H}$^0$}}

\DeclareMathAlphabet{\vib}{OML}{cmm}{m}{it}

\def\h{$H_{\rm 160}$}

\def\ics{IRAC-selected sources}

\begin{document}
\title[Examining why RLAGN are located in proto-clusters]{Why $z>1$ radio-loud galaxies are commonly located in proto-clusters}
\author[N.\,A.\,Hatch et al.]
       {\parbox[]{6.0in}
       {N.\,A.\,Hatch$^{1}$\thanks{E-mail: nina.hatch@nottingham.ac.uk},  D.\,Wylezalek$^2$,  J.\,D.\,Kurk$^{3}$, D.\,Stern$^4$, C.\,De\,Breuck$^2$, M.\,J.\,Jarvis$^{5,6}$,~A.\,Galametz$^{3,7}$,~A.\,H.\,Gonzalez$^8$,~W.\,G.\,Hartley$^9$,~A.\,Mortlock$^{10}$, N.\,Seymour$^{11,12}$, J.\,A.\,Stevens$^{13}$\\        \footnotesize
        $^1$School of Physics and Astronomy, University of Nottingham, University Park, Nottingham NG7 2RD\\  
	$^2$European Southern Observatory, Karl-Schwarzschild-Str. 2, D-85748 Garching, Germany\\
	   $^3$Max-Planck-Institut fuer Extraterrestrische Physik, Giessenbachstrasse, D-85748 Garching, Germany\\  	
	$^4$Jet Propulsion Laboratory, California Institute of Technology, 4800 Oak Grove Dr., Pasadena, CA 91109, USA\\
	$^5$Astrophysics, Department of Physics, Keble Road, Oxford OX1 3RH\\
	$^6$Physics Department, University of the Western Cape, Bellville 7535, South Africa\\
	$^7$INAF - Osservatorio di Roma, Via Frascati 33, I-00040, Monteporzio, Italy\\
	$^8$Department of Astronomy, University of Florida, Gainesville, FL 32611, USA\\
	$^9$ETH Z\"urich, Institut f\"ur Astronomie, HIT J 11.3, Wolfgang-Pauli-Str. 27, 8093, Z\"urich\\
	$^{10}$Institute for Astronomy, University of Edinburgh, Royal Observatory, Edinburgh EH9 3HJ\\
	$^{11}$CASS, PO Box 76, Epping, NSW, 1710, Australia\\
	$^{12}$Curtin Institute of Radio Astronomy, Curtin University, GPO Box U1987, Perth WA 6845, Australia\\
	$^{13}$Centre for Astrophysics Research, STRI, University of Hertfordshire, Hatfield AL10 9AB    
    }}
 \date{}
\pubyear{}
\maketitle

\label{firstpage}
\begin{abstract}
Distant powerful radio-loud active galactic nuclei (RLAGN) tend to reside in dense environments and are commonly found in proto-clusters at $z>1.3$. We examine whether this occurs because RLAGN are hosted by massive galaxies which preferentially reside in rich environments. We compare the environments of powerful RLAGN at $1.3<z<3.2$ from the CARLA survey to a sample of radio-quiet galaxies matched in mass and redshift. We find the environments of RLAGN are significantly denser than those of radio-quiet galaxies, implying that not more than 50\%  of massive galaxies in this epoch can host  powerful radio-loud jets. This is not an observational selection effect as we find no evidence to suggest it is easier to observe the radio emission when the galaxy resides in a dense environment. We therefore suggest that the dense Mpc-scale environment fosters the {\em formation} of a radio-jet from an AGN. We show that the number density of potential RLAGN host galaxies is consistent with every $>10^{14}$\Msun\ cluster having experienced powerful radio-loud feedback of duration $\sim$60\,Myr during $1.3<z<3.2$. This feedback could heat the intracluster medium to the extent of 0.5-1\,keV per gas particle, which could limit the amount of gas available for further star formation in the proto-cluster galaxies.

\end{abstract}
\section{Introduction}
Radio-loud Active Galactic Nuclei (RLAGN) are typically located in dense environments \citep[e.g.,][]{Yates1989, Hill1991, Roche1998, Best1998, Best2000, Donoso2010}. At $z\gtsim1.5$, many of these  regions are dense enough that they will collapse into clusters by today, so they are commonly referred to as proto-clusters \citep[e.g.,][]{Venemans2007,Hatch2011a}. 

The Clusters Around Radio-Loud AGN  survey (CARLA) recently showed that approximately half of all powerful RLAGN (L$_{\rm 500MHz}\ge10^{27.5}$\,W\,Hz$^{-1}$) at $1.3<z<3.2$ reside in regions that are denser than average by more than $2\sigma$. Many of these dense regions are likely to be proto-clusters \citep{Wylezalek2013,WylezalekLF}.  Low-luminosity RLAGN (L$_{\rm 1.4GHz}\sim10^{25.5}$\,W\,Hz$^{-1}$) also tend to reside in rich groups and clusters. \citet{Castignani2014} showed that $\sim70$\% of low-luminosity RLAGN at $1<z<2$ are surrounded by Mpc-scale galaxy overdensities.  \citet{Castignani2014} also demonstrated that the simple cluster-detection method used by the CARLA survey \citep{Wylezalek2013,WylezalekLF} missed many of the clusters around the low-luminosity RLAGN, so it is likely that the fraction of powerful RLAGN that reside in proto-clusters is higher than 50\%. 
 These surveys prove that RLAGN efficiently trace galaxy proto-clusters; the goal of this paper is to understand why RLAGN are such good beacons of proto-clusters.

At $z\le0.7$ RLAGN occupy richer environments than similarly massive radio-quiet galaxies \citep[e.g.,][]{Kauffmann2008, Ramos_Almeida2013}. This implies that the presence of a radio-jet depends on environment as well as galaxy mass. However, this may not be true at higher redshifts.  Powerful RLAGN are hosted by galaxies with a stellar mass of  $>10^{10.5}$\Msun\ \citep{Seymour2007}, and the bias of such massive galaxies implies they reside in dark matter halos of $10^{12.5}$\Msun\ or greater at all redshifts \citep{Hartley2013}. In the local Universe, this means most massive galaxies reside in group environments, but such massive halos at $z>1.5$ typically grow into cluster-mass structures by today \citep{Chiang2013}. Thus massive galaxies at high-redshift are likely to trace the progenitors of rich cluster environments, and the spatial correspondence between RLAGN and proto-clusters may simply occur because distant RLAGN are hosted by massive galaxies. In this paper we compare the environments of $1.3<z<3.2$ RLAGN from the CARLA survey to similarly massive galaxies without radio-jets, to determine whether RLAGN reside in proto-clusters simply  because they are massive galaxies, or if the presence of a radio-jet depends on environment at high-redshift, as it does in the local Universe.

Another reason why powerful RLAGN may be such good tracers of proto-clusters is that the radio-emission could be amplified if the relativistic electrons are constricted by the dense ambient gas \citep{Barthel1996}. An enhancement of the radio power would produce a selection bias, meaning it is easier to observe RLAGN when they reside in dense environments. Several studies have investigated whether this selection bias exists by searching for a correlation between radio-power and environmental density, but the results are contradictory. For example, \citet{Karouzos2014} found no trend, \citet{Donoso2010} found a negative trend, and \citet{Falder2010} found a positive trend between radio power and environmental density. \citet{Wylezalek2013} showed there is no correlation between the radio power of the CARLA AGN and their environment, suggesting that no selection bias exists at $z>1.3$. Nevertheless, there is so much confusion in the literature on this topic that a more in-depth  analysis is warranted. Therefore we will perform a more detailed comparison of the radio properties of the CARLA AGN with environment to determine if the ambient gas affects the radio emission.

The purpose of this study is to explore why RLAGN are such good beacons for locating proto-clusters. We test two hypothesises: (i) that RLAGN are in denser environments simply because they are hosted by massive galaxies,  and (ii) that the dense proto-cluster environment amplifies the radio emission causing a selection bias. In Section\,\ref{method} we introduce our data and describe how we create a radio-quiet control sample that is matched in redshift and mass proxy to the CARLA RLAGN sample. In Section \ref{results} we compare the environments of the RLAGN and radio-quiet massive galaxies, then search for any correlation between the properties of the radio-emission and the surrounding environment. We discuss the implications of our results in Section \ref{discussion}. We use AB magnitudes throughout and a $\Lambda$CDM flat cosmology with $\Omega_M=0.3$, $\Omega_\Lambda=0.7$ and $H_0=70$\kmpspMpc.

\section{Method}
\label{method}
\subsection{Data}
\subsubsection{RLAGN sample:  CARLA}
CARLA is a $400$-hour Warm {\it Spitzer Space Telescope} programme designed to investigate the environments of powerful RLAGN.  The CARLA sample consists of 419 very powerful RLAGN lying at $1.3<z<3.2$ and having a 500\,MHz luminosity $\ge10^{27.5}$\,W\,Hz$^{-1}$. The sample comprises of 211 radio-loud quasars (RLQs) and 208 radio galaxies (RGs). 

The RGs at $z>2$ were selected from the compendium of \citet{MileyDeBreuck2008}; RGs at $z < 2$ were selected from flux-limited and ultra-steep spectrum radio surveys to have the same distribution of radio-power as the higher-redshift sample. The RLQ sample comprises of optically bright ($M_B<-26.5$) quasars in SDSS \citep{Schneider2010} and the 2dF QSO Redshift Survey \citep{Croom2004} with NVSS detections above the L$_{\rm 500MHz}\ge10^{27.5}$\,W\,Hz$^{-1}$ threshold. More RLQs than HzRGs matched these criteria, so we limited the RLQ sample to 211 sources which matched the redshift and radio-power distribution of the HzRG sample. Full details of the CARLA RLAGN sample can be found in \citet{Wylezalek2013}. No information about the environments was taken into account when selecting targets, so the sample is representative of the entire powerful RLAGN population at these redshifts.

Deep {\it Spitzer} data covering $5.2$\arcmin$\times 5.2$\arcmin\ were obtained on each field with the Infrared Array Camera (IRAC; \citealt{Fazio2004}) on board the {\it Spitzer Space Telescope} at 3.6 and 4.5 $\micron$ during cycles 7 and 8. The CARLA  95\% completeness limiting magnitudes are [3.6] = 22.6 and [4.5] = 22.9\,mag. Details of the observations and data reduction can be found in \citet{Wylezalek2013}. Using the well-tested {\it Spitzer} IRAC colour criterion $[3.6]-[4.5] > -0.1$ to select galaxies at $z > 1.3$, \citet{Wylezalek2013} showed that 55\% of these RLAGN are surrounded by significant excesses of galaxies that are likely associated with the RLAGN.

\begin{figure*}
\includegraphics[width=2\columnwidth]{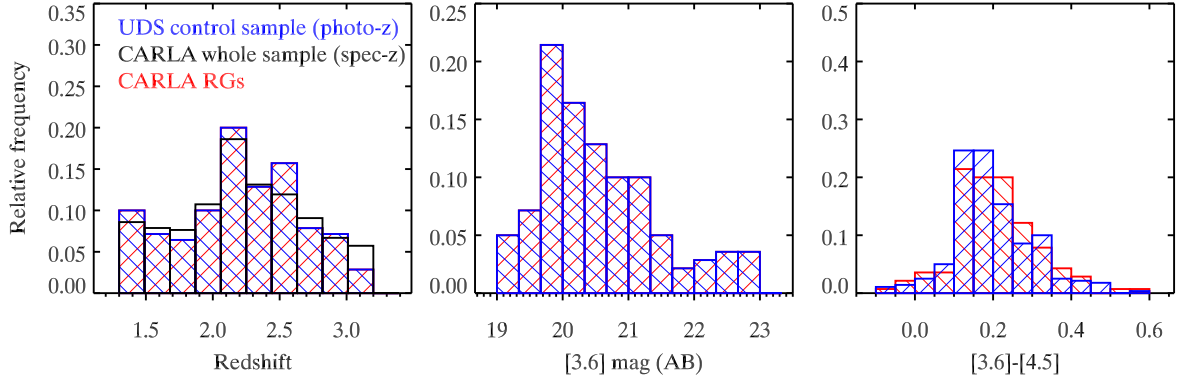}
\caption{The distributions of the redshifts (left), [3.6] magnitudes (middle), and [3.6]-[4.5] colours (right) of the CARLA reduced radio galaxy sample (red), the UDS control sample (blue) and the whole CARLA sample (black; shown in the left panel only as the IRAC magnitudes and colours of the RLQs are dominated by the quasar light. See Section\,\ref{matched_catalogue} for details). The control sample is closely matched to the CARLA radio galaxy sample; KS tests result in probabilities in the range of 0.3-0.97, so there are no significant differences between these distributions. \label{match}}
\end{figure*}
\subsubsection{Control fields: UDS and SpUDS}
To obtain a control field sample of radio-quiet galaxies with the same mass and redshift distribution as the RLAGN, we utilise the UKIDSS Ultra Deep Survey (UDS; Almaini et al., in prep.). The UDS is a deep 0.8\,deg$^2$ near-infrared survey overlapping part of the {\it Subaru}/{\it XMM-Newton} Deep Survey (SXDS; \citealt{Furusawa2008}). In the overlapping $\sim$0.54\,deg$^2$ area the total co-moving volume between $z=1.3$ and $3.2$ is 0.012\,Gpc$^{-3}$. In this volume we expect  approximately 50 proto-clusters that will collapse to form clusters of M$_{200}\ge10^{14}$\Msun\  (assuming local galaxy cluster number counts from \citealt{Vikhlinin2009}).

A sample of massive galaxies are identified from the UDS using the photometric redshifts and stellar masses derived by \citet{Hartley2013} and \citet{Mortlock2013}. These authors combine $U-$band data from the Canada-France-Hawaii Telescope (Foucaud et al., in preparation) with optical photometry from the SXDS, {\it JHK} photometry from the 8th data release (DR8) of the UDS, and  {\it Spitzer} Ultra Deep Survey data (SpUDS; PI J. Dunlop) to create a $K-$selected $UBVRizJHK$[3.6][4.5] catalogue. 

Photometric redshifts were determined by fitting spectral energy distribution (SED) templates to the photometric data points. Following \citet{Hartley2013} we remove objects with poorly determined photometric redshifts by removing the objects whose minimum $\chi^2$ is greater than $11.35$ from  the photometric redshift fitting procedure (15\% of the sample). Many of the removed objects are blended sources or optically bright AGN, so the control sample may be biased against galaxies containing optically bright AGN. These objects have average environments, so removing them from the UDS catalogues does not bias our results. The dispersion of the remaining photometric redshifts is $\Delta z$/($1 + z$)$ = 0.031$. For full details regarding the methodology and resulting photometric redshifts of the UDS catalogue, see \citet{Hartley2013}.

The stellar masses of the $K-$selected UDS galaxies were measured by fitting the photometry to a large grid of synthetic SEDs from \citet{BC03} stellar population models (created using a Chabrier initial mass function; \citealt{Chabrier2003}). Full details of the fitting procedure are given in \citet{Mortlock2013}. The catalogue is 95 per cent complete to log\,$ (M/M_{\odot})=10.3$ at the highest redshift of the CARLA sample ($z=3.2$).  

The UDS galaxies were classified as star-forming or quiescent using the two colour rest-frame  $UVJ$ selection introduced by Wuyts et al. (2007). Our division follows the boundaries defined by Williams et al. (2009), and extends their $1<z<2$ values to higher redshift. However, quiescent galaxies selected by this method can be contaminated by AGN and dusty star-forming objects. Therefore we reclassify galaxies as star-forming if they are fit by a template with sSFR $>10^{-8}$\,yr$^{-1}$, or if the galaxy is associated with a {\it Spitzer} $24\micron$ source, with a flux that would imply a sSFR $>7.43\time10^{-11}$\,yr$^{-1}$ (a stellar mass doubling time less than the $z=0$ Hubble time). As shown by \citet{Stern2006}, the $24\micron$ data reliably breaks the degeneracy between passive and dusty red galaxies.

The environment of the UDS massive galaxies were measured using SpUDS, a 1\,deg$^2$ cycle 4 {\it Spitzer} Legacy program which encompasses the UDS field. We used the SpUDS $3.6$ and $4.5\micron$ catalogues of \citet{Wylezalek2013}, which were extracted from the public  mosaics in the same way as for CARLA. The catalogues were extracted in dual-image mode with the $4.5\micron$ image used as the detection image. The SpUDS data reach 3$\sigma$ sensitivities of $[3.6]\sim[4.5]\sim24$\,mag, but in all following work the catalogues were cut to the shallower depth of the CARLA data.

\subsection{Control sample}

\subsubsection{Creating a control sample matched in mass and redshift to CARLA RLAGN}
\label{matched_catalogue}
In this section we describe how a radio-quiet control sample of galaxies, with similar masses and redshifts as the CARLA sample, were selected from the UDS catalogues. We first removed all sources within 2\,arcmin of bad regions of the SpUDS data, so that we do not include sources whose environments are affected by bright stars or field edges. 

To form a radio-quiet sample we removed all UDS galaxies from the catalogue that were detected in the 100$\mu$Jy {\it Subaru/XMM-Newton} Deep Field radio source sample of \citet{Simpson2006}. There are 109 sources with radio fluxes greater than 100$\mu$Jy in the redshift, [3.6] magnitude and colour range of the CARLA sample, most of which lie at $z<1.8$. Although we refer to the remaining galaxy sample as \lq radio-quiet\rq\ we note that the 100$\mu$Jy flux density limit of the  \citet{Simpson2006} catalogue means that the sample may still include radio-emitting galaxies with a 500\,MHz luminosity of $10^{24.4}$W\,Hz$^{-1}$ at $z=1.3$ and up to $10^{25.2}$W\,Hz$^{-1}$ at $z=3.2$ (assuming a spectral index $\alpha =-0.9$: see Fig.\,\ref{fig:radio_prop}c). These radio luminosities are at least two orders of magnitude lower than the radio luminosities of the CARLA RLAGN.

The next step was to select galaxies which have a similar distribution of stellar mass as the CARLA  RLAGN. Radio-loud galaxies are among the most massive galaxies at every redshift; \citet{Seymour2007} found that none of the 70 radio galaxies in the {\it Spitzer} high-redshift radio galaxy (SHzRG) programme had a stellar mass\footnote{The SHzRG galaxy masses were reduced by 13\% to account for the differences in assumed initial mass functions in the SED fitting (\citealt{Kroupa2001} for SHzRG versus \citealt{Chabrier2003} for the UDS).} $<10^{10.3}$\Msun, so all UDS galaxies with stellar masses $<10^{10.3}$\Msun\ were removed from the control sample. We tested the impact of this mass cut by leaving in all control galaxies with $<10^{10.3}$\Msun, and find that our results do not change.

To derive the stellar masses of the CARLA RLAGN requires multicomponent SED fitting that takes into account light from the AGN as well as the stellar population. To do this requires mid-infrared data at wavelengths $>5$\micron\ to remove the hot dust component \citep[e.g.][]{DeBreuck2010,Drouart2014}, and high-resolution data to remove the direct AGN light \citep[e.g.][]{Hatch2013}. Since these data are not available for the majority of the CARLA sample we are unable to perform adequate fits to the SEDs to obtain stellar masses. 

An alternative to deriving stellar masses is using the {\it Spitzer} IRAC fluxes and colours as mass proxies. The {\it Spitzer} [3.6] and [4.5] bands cover rest-frame near-infrared emission ($0.8-2$\micron)  for galaxies at $1.3<z<3.2$, hence are relatively good tracers of stellar mass. We cannot use the {\it Spitzer} IRAC fluxes of the CARLA quasars as mass proxies, as these objects are dominated by the light from the central AGN. However, the IRAC fluxes of the 208 CARLA radio galaxies, whose central AGN light is mostly obscured, are likely to be good mass proxies.  

To form the control sample we divided the 208 radio galaxies in the CARLA sample into ten redshift bins of width $\Delta z=0.19$; all CARLA RLAGN have spectroscopically measured redshifts. Each redshift bin was subdivided into twelve [3.6] magnitude bins of $\Delta~m=0.33$, which was further subdivided into four bins of [3.6]-[4.5] colour with $\Delta$\,colour\,$=0.3$. Thus the 208 CARLA radio galaxies were divided into 480 bins. The radio-quiet UDS galaxies were divided into the same 480 bins, and from each bin we randomly selected twice the number of UDS galaxies as CARLA radio galaxies.  Eighteen bins contained fewer UDS galaxies than CARLA radio galaxies, therefore we gave an extra weight to the 38 UDS galaxies in these bins so they had the equivalent weight of 114 galaxies. The final control sample consisted of 280 galaxies.

A subset of 68 CARLA radio galaxies fell in bins that were not occupied by any UDS source. These sources were generally bright and red galaxies at high-redshift. It is likely that the IRAC fluxes from these galaxies are contaminated by direct AGN light or hot dust.  The caveat to using IRAC magnitudes as a mass proxy is that approximately a third of radio galaxies at $1.3<z<3.2$ are dominated, at rest-frame 1.6\micron, by AGN-heated hot dust emission rather than stellar emission  \citep{DeBreuck2010}. These galaxies appear slightly brighter and redder than galaxies of similar mass without AGN.  We thus removed the 68 galaxies from the CARLA radio galaxy sample to leave a subsample of 140 galaxies that matches the control sample perfectly in redshift, [3.6] and [3.6]-[4.5]. Removing these galaxies does not affect our results as the redshifts and environmental densities of the remaining CARLA radio galaxies are very similar to the full CARLA sample (as shown in Figs.\,\ref{match} and \ref{main}). It is possible that some RLAGN in our remaining CARLA radio galaxy sample are still contaminated by low levels of hot dust emission in the IRAC bands, however, this means that our mass-matching will be on the conservative side, and the control galaxies selected may be slightly more massive than the CARLA radio galaxies.

In Fig.\,\ref{match} we compare the redshifts, [3.6], and [3.6]-[4.5] colours of the CARLA radio galaxy subsample and the control sample. Kolmogorov-Smirnov (KS) tests result in probabilities of 0.77, 0.97, and 0.33, for the redshifts, magnitudes and colours, respectively. These tests suggest there are no significant differences in these properties between the CARLA radio galaxy subsample and UDS control galaxies.

The fraction of control galaxies that host radio-quiet AGN was measured using deep {\it XMM-Newton} data \citep{Ueda2008}. The closest $K$-band selected galaxy within 5\,arcsec of the X-ray point source was assumed to be the galaxy counterpart, which resulted in 191 X-ray point sources in the redshift range $1.3<z<3.2$. Approximately 7\%  of the control galaxies have X-ray detections indicating they host radio-quiet AGN. There is no significant difference in the IRAC luminosity, colour or environment between the X-ray-bright control galaxies and the rest of the control sample. 

\subsubsection{Stellar masses of the CARLA and control galaxies}
\label{mass_test}     

One of the aims of this work is to test whether RLAGN reside in denser environments than similarly  massive radio-quiet galaxies. Hence the validity of our results hinges on our selection of a control sample that is well-matched in both redshift and mass. The redshift selection is trivial since the CARLA RLAGN all have spectroscopic redshifts and the photometric redshifts of the UDS catalogue are of high quality \citep{Hartley2013}. Ensuring that our control sample matches the stellar masses of the CARLA RLAGN is harder because we must rely on mass proxies.

\begin{figure}
\includegraphics[width=1\columnwidth]{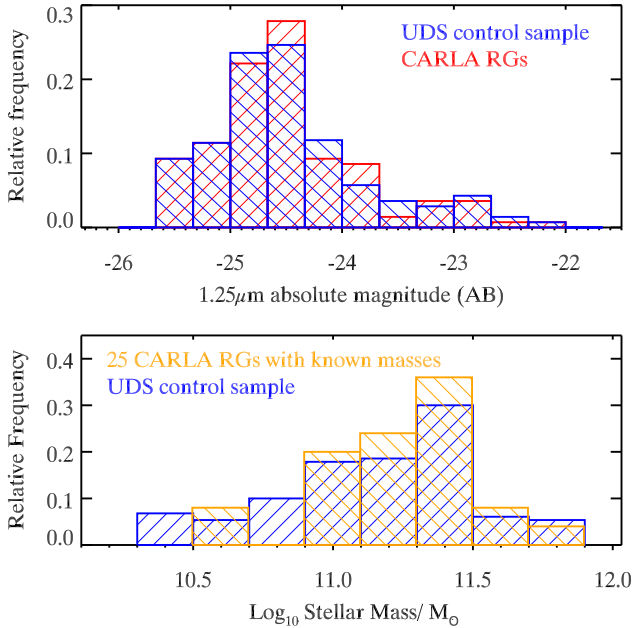}
\caption{A comparison of the rest-frame 1.25$\micron$ absolute magnitudes (top panel) and stellar masses (bottom panel) between the CARLA RLAGN and the control galaxies. There is no significant difference in the rest-frame 1.25$\micron$ absolute magnitudes, and the stellar masses of the control galaxies are very similar to the stellar masses of the 25 CARLA RLAGN that were measured using SED fitting of $3.6$ to $70\micron$ photometry by \citet{Seymour2007}. Hence the control galaxies are well-matched in stellar mass to the CARLA RLAGN. \label{match_mass}}
\end{figure}

The near-IR luminosity of a galaxy is a good measure of its stellar mass \citep{Kauffmann1998}, therefore we convert the IRAC [3.6] and [4.5] photometry of the RLAGN and control galaxies to rest-frame luminosities using the spectroscopic and photometric redshifts, respectively. We then linearly interpolate or extrapolate these two data points to measure the rest-frame 1.25$\micron$ luminosities of the galaxies. This wavelength is close to the $J-$band central wavelength and was chosen because it lies in between the observed [3.6] and [4.5] bands for the majority of the galaxies.  As illustrated in the top panel of Fig.\,\ref{match_mass}, the CARLA radio galaxies and control galaxies are so similar in rest-frame 1.25$\micron$ absolute magnitude that a KS test is unable to distinguish between these samples (P=0.97).

Approximately a fifth of our reduced CARLA radio-galaxy subsample have stellar masses measured by  \citet{Seymour2007} using SED fitting of $3.6$ to $70\micron$ photometry. These masses are very similar to those of the control galaxies (measured through SED fitting of $U-$band to $4.5\micron$ photometry), as shown in the bottom panel of Fig.\,\ref{match_mass}. A KS test results in P=0.15, which means there is no significant difference between the masses of these 25 RLAGN and the control galaxies. These two tests suggest that the control galaxy sample is well-matched in mass to the CARLA RLAGN. 

\subsubsection{Galaxy type}
Whilst forming a control sample it is important to consider galaxy type, as passive galaxies are located in denser environments than their star-forming counterparts up to $z\sim1.8$ \citep{Chuter2011,Quadri2012}. RLAGN at $1.3<z<3.2$ do not conform to a single galaxy type.  RLAGN hosts at $z\lsim1.8$ tend to contain old, quiescent stellar populations, and their light profiles are similar to the de Vaucouleurs profile \citep[e.g.,][]{Best1998}. In contrast, RLAGN hosts at $z\gsim2$ often have clumpy morphologies and high star formation rates \citep{Pentericci1999,Drouart2014}. It is likely that the CARLA RLAGN comprise a range of galaxy types. The control sample similarly contains a mix of galaxy types, with approximately 55\% defined as quiescent. 

With only IRAC colour information for the CARLA RLAGN it is not possible to form a control sample that matches galaxy type exactly. However, we checked our results whilst limiting the control sample to quiescent galaxies only, and then with star-forming galaxies only, and found no difference in the results (see Section \ref{section3.1} for more details).

\subsection{Measuring environment}
The environments of the CARLA RLAGN and control galaxies were measured by counting the number of {\it Spitzer} IRAC colour-selected galaxies within 1\arcmin\ radius circles centred on each galaxy in the samples. This radius was chosen because it corresponds to an angular size of $\sim0.5$\,Mpc at $1.3<z<3.2$, which is the typical radius of a high-redshift cluster. The well-tested $[3.6]-[4.5]>-0.1$ colour-cut was used to remove galaxies with $z<1.3$ (\citealt{Papovich2008,Muzzin2013a,Rettura2014}, see discussion in \citealt{Galametz2012}). Using this criterion we expect to obtain a $z>1.3$ galaxy sample with only $10-20$\% contamination by low-redshift interlopers \citep{Muzzin2013a}. This colour selection criterion identifies a homogenous sample of galaxies out to $z\sim3.2$ because  [4.5] is almost constant for galaxies at $z>0.7$ due to a negative $k$-correction, and these wavelengths cover the 1.6$\micron$ stellar bump for galaxies at $1.3<z<3.2$. This stellar bump is a prominent feature regardless of a galaxy's star formation history. For the rest of this work we refer to the objects matching the $[3.6]-[4.5]>-0.1$ colour selection as \lq\ics\rq.

An object is defined as an IRAC-selected source if it is detected above the $4.5\micron$ 95\% completeness limit of the CARLA data ([4.5]$=$22.9\,mag) and has a colour of $[3.6]-[4.5] > -0.1$. If the source is not detected at $3.6\micron$, an upper limit of the [3.6]-[4.5] colour is determined using the $3.5\sigma$ detection limit of the $3.6\micron$ CARLA data ([3.6]=22.8\,mag). This is the same criteria used as in \citet{WylezalekLF}, although differs from that used in \citet{Wylezalek2013} in terms of the $3.6\micron$ depth.

Most galaxies associated with the RLAGN will have magnitudes fainter than $m^{*}-1$, where $m^{*}$ is the characteristic apparent magnitude of the proto-clusters surrounding the RLAGN \citep{McLure2001}. \citet{WylezalekLF} measured $m^{*}$ at [4.5] for all proto-clusters associated with the CARLA RLAGN in 6 redshift bins spanning $1.3<z<3.2$. They found $m^{*}$ to be in the range of 19.85 to 20.41$\pm{0.20}$\,mag, which is consistent with the proto-cluster galaxies forming at $z\sim3$ and passively evolving thereafter. $m^{*}$ at 4.5$\micron$ is approximately constant across $1.3<z<3.2$ because of a negative $k-$correction caused by the $1.6\micron$ stellar bump that enters the {\em Spitzer} IRAC bands in this redshift range. Therefore, we assume the average  $m^{*}$ of $\sim20.1$\,mag is valid for all the CARLA proto-clusters\footnote{We found no difference in our results if we used the full range of $m^{*}$ found by \citet{WylezalekLF} as apposed to the average $m^{*}$.}, and any IRAC-selected source with [4.5]$<19.1$\,mag was not included in our measurement of the environment for both the RLAGN fields and the control field. This selection should also remove many of the low-redshift interlopers.

\subsection{Properties of CARLA RLAGN}
If RLAGN are good tracers of proto-clusters because they are easier to detect when they reside in dense environments, we expect to see correlations between their environment and their radio-properties.  We therefore measured the spectral index, $\alpha$ (where $S_\nu\propto\nu^\alpha$), bolometric luminosity, radio emission extent and supermassive black hole (SMBH) mass of the CARLA RLAGN to investigate possible trends with our environmental measurement. 

\subsubsection{Black hole masses and bolometric quasar luminosity}
The SMBH masses and the bolometric luminosities of the 211 radio-loud quasars in the CARLA sample were  obtained from the SDSS  \citep{Shen2011}. The SMBH masses are virial masses, i.e.,\,it is assumed that the broad-line emission region is virialised and the continuum luminosity and broad emission line width are good proxies for the broad-line region radius and velocity, respectively. Due to the spectral coverage of the SDSS data the Mg{\sc ii} line is used to calculate the virial mass of the SMBH for RLAGN at $z<1.9$ , whilst for all higher redshift RLAGN the C{\sc iv} line is used \citep{McLure2002,Vestergaard2002}. We note that there are large uncertainties and significant systematic biases associated with these SMBH mass estimates, which are described in detail in \citet{Shen2011}.

\subsubsection{Extent of the radio emission}
The sizes of the radio emission of CARLA sources were measured from the Very Large Array (VLA)  Faint Images of the Radio Sky at Twenty-cm (FIRST) survey \citep{White1997} at $\sim1.4$\,GHz which has a resolution of 5\arcsec. We matched the 2013 June 05 version of the FIRST catalogue with the CARLA sources and identified 284 sources covered by FIRST. All sources in the catalogue that have a $>5$\% probability of being a spurious source were removed; the RLAGN are very bright so it is common that nearby sources are sidelobes. 

We identified RLAGN which had multiple radio components within 1\arcmin, and defined the radio extent as the largest distance between the radio sources detected within 1\arcmin\ of the RLAGN host galaxy. All sources were visually inspected for classical radio component distribution. For AGN with only single radio components we identified the closest radio counterpart within 5\arcsec\ and the size was defined as the full-width-at-half-maximum (FWHM) of the major axis, which had been deconvolved to remove blurring by the elliptical Gaussian point-spread function, down to a major axis FWHM $<2$\arcsec. Sources with major axis FWHM $<2$\arcsec\ were classified as unresolved. Of the 284 CARLA RLAGN with FIRST coverage, 127 are extended sources (74 with multiple radio components) and 157 are unresolved sources. 
\subsubsection{Spectral indices}
The spectral index of the radio emission was measured by cross-correlating two radio surveys with similar spatial resolution (45-80\arcsec): the 1.4\,GHz NRAO VLA Sky Survey  \citep{Condon1998} and the 74\,MHz VLA Low-Frequency Sky Survey \citep{Cohen2007}. 
 Radio-loud quasars may emit time-variable Doppler beamed emission so it is not reliable to measure their spectral indices from surveys taken several years apart. Therefore we only measure the spectral indices for the 158 radio-loud galaxies in the CARLA sample that are covered by both the NVSS and VLSS radio surveys.

\section{Results}
\label{results}
\begin{figure}
\includegraphics[width=1\columnwidth]{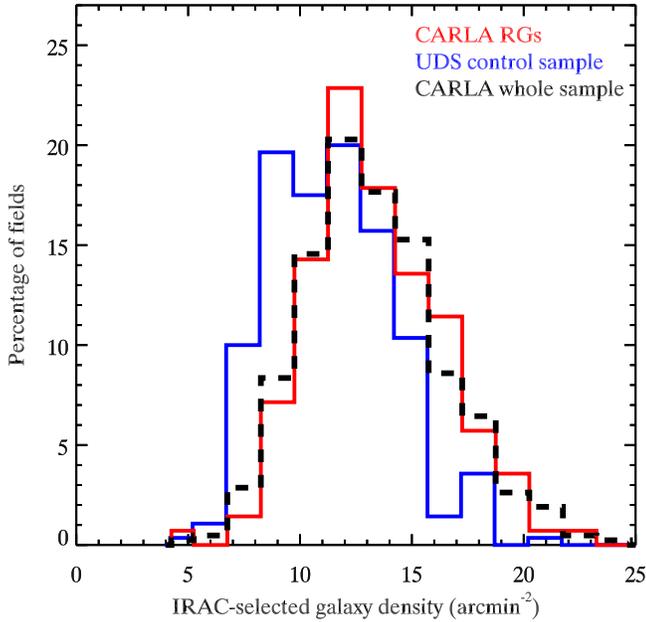}
\caption{A comparison of the environment surrounding the CARLA RLAGN (red and black) and the UDS control  sample (blue). The environment is assessed using the density of \ics\ within an arcmin radius. The black dashed histogram shows the full CARLA sample. The solid red histogram comprises of the radio galaxy CARLA subsample described in Section \ref{matched_catalogue}. The environments of the subsample and full CARLA sample are statistically indistinguishable (KS test results in $P=0.95$). However, the environment of the control sample differs significantly from both CARLA samples (KS $P\sim10^{-4}$).  \label{main}}
\end{figure}

\subsection{Do RLAGN trace proto-clusters simply because they are hosted by massive galaxies?}
\label{section3.1}

If RLAGN reside in dense environments simply because they are hosted by massive galaxies, then we expect massive radio-quiet galaxies to occupy similarly dense environments.  In Fig.\,\ref{main} we compare the environments of CARLA RLAGN to the control sample. The environments of all 419 RLAGN in the CARLA sample are very similar to the radio-galaxy subsample (a KS test gives P=0.95). The environments of the control galaxies are on average less dense than the surroundings of the RLAGN in both CARLA samples. A KS test results in a probability of $<10^{-4}$ ($\sim4\sigma$ significance) that the radio-loud and radio-quiet galaxies reside in similar environments. This means the high mass of the RLAGN hosts is not the only reason why they trace rich environments. 

Galaxy type may also influence our results as quiescent galaxies are located in denser environments than actively star-forming galaxies, at fixed stellar mass, even up to $z\sim1.8$ \citep{Quadri2012}. We therefore checked the influence of galaxy type on our results by limiting the control sample to quiescent galaxies only, and then with star-forming galaxies only. We found no difference in the results, with both quiescent and star-forming control samples differing from the CARLA sample with $4\sigma$ significance. This means that RLAGN reside in richer environments than all types of similarly massive galaxies.

\begin{figure}
\centering
\includegraphics[width=1.\columnwidth]{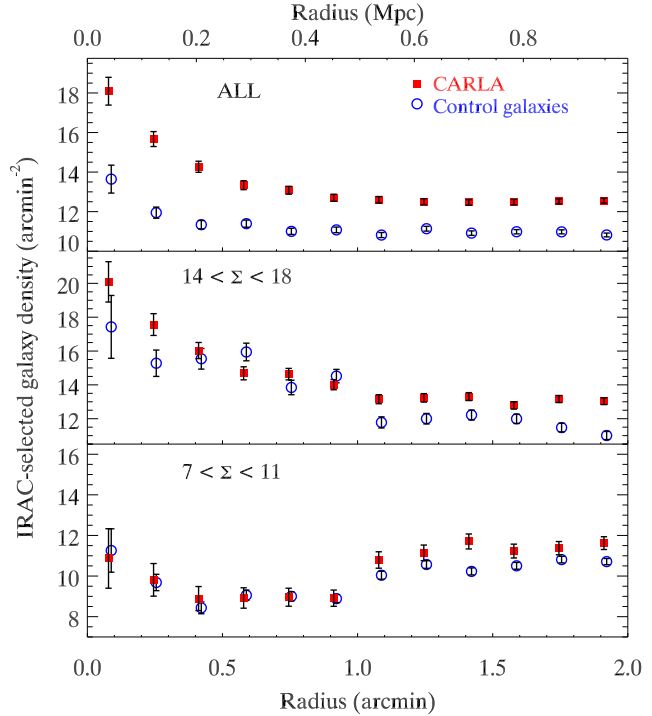}
\caption{The radial profile of the IRAC-selected galaxy density surrounding CARLA RLAGN and control galaxies. The top panel displays all CARLA and control galaxies, the middle panel shows galaxies  which have central arcmin densities between 14 and 18 IRAC-selected  galaxies per arcmin$^{2}$, and the bottom panel shows galaxies with central arcmin densities in the range of 7 to 11 IRAC-selected galaxies per arcmin$^{2}$. The central RLAGN and massive control galaxy are not included, and uncertainties are $\sqrt{N}$. The RLAGN are in denser environments on all scales. In the bottom two panels the profiles of the central arcmin are similar (by construction), but at larger radii the RLAGN are denser than the control galaxies indicating that larger and more massive structures surround the RLAGN. The radius is shown both in arcmin, and proper angular distance for galaxies at $z=2.2$ (which is the median redshift of the CARLA sample).  \label{fig:radial_profile}}
\end{figure}

In Fig.\,\ref{fig:radial_profile} we show the difference between the environments of the RLAGN  and the control galaxies in more detail by comparing their average radial density profiles. The top panel of Fig.\,\ref{fig:radial_profile} shows that the environments of RLAGN are denser than the control galaxies at all scales, having both more nearby neighbours and a larger excess beyond 0.5\,Mpc.

We then compare the radial profiles surrounding RLAGN and control galaxies that have similar densities within a 1\,arcmin radius, choosing both high density regions with $14<\Sigma<18$ IRAC-selected galaxies per arcmin$^{2}$ (middle panel of Fig.\,\ref{fig:radial_profile}) and  underdense regions with $7<\Sigma<11$ IRAC-selected galaxies per arcmin$^{2}$ (bottom panel of Fig.\,\ref{fig:radial_profile}). The central densities are comparable by construction, but beyond 1\,arcmin ($\sim0.5$\,Mpc) the profiles diverge: RLAGN reside in denser large-scale environments than the control galaxies.

This excess at large radii is not an artefact of the data reduction as the raw number counts of the SpUDS control field and the CARLA fields are in good agreement (shown in Fig.\,2 of \citealt{Wylezalek2013}). These radial profiles imply the structures surrounding the RLAGN are more extended than those around massive radio-quiet galaxies. It is possible that the structures surrounding RLAGN are more massive, or alternatively, the extended structure may be the signature of merging galaxy groups or clusters.  \citet{Simpson2002} suggested that powerful RLAGN, like the CARLA galaxies, may pinpoint merging clusters as they may be triggered by galaxy-galaxy interactions that occur during mergers.  Regardless of the origin of the extended structure, Fig.\,\ref{fig:radial_profile} shows that RLAGN are more likely to reside in high-mass groups and clusters than the average radio-quiet massive galaxy, even if their  $<1$\,arcmin environment (0.5\,Mpc) appears average or under-dense.

Overall, these results mean that the probability a radio-jet is launched from an AGN depends on the Mpc-scale environment, as well as the galaxy mass.  These results are in agreement with studies of $z<0.7$ RLAGN, such as \citet{Kauffmann2008} and  \citet{Ramos_Almeida2013} who showed that RLAGN are located in significantly denser environments than similarly massive samples of quiescent galaxies. They also suggest that the high density environment promotes the launch of radio jets from SMBHs. Our results extend these studies to the high-redshift Universe, and show that environment continues to have a strong influence on AGN properties at $1.3<z<3.2$.

The overdensities surrounding the CARLA RLAGN extend beyond 2\arcmin, i.e.\,$2.3-4.2$\,cMpc  (co-moving Mpc). According to the models of \citet{Chiang2013}, this implies that the typical masses of the overdensities are greater than $10^{14}$\Msun. This mass is spread across a few hundred  cMpc$^{3}$, and many of these regions are likely to be diffuse proto-clusters rather than rich clusters. In time, these regions will collapse to become rich clusters, but at the epoch of observation, the local environment of the high-redshift RLAGN may be similar to that of low- and intermediate-redshift RLAGN, which are generally located in galaxy groups and poor clusters \citep{Fisher1996,Best2000,Best2004,McLure2001}. It is possible that this particular environment, where galaxy mergers and harassment are likely to be frequent, is very efficient at creating radio-jets.

\begin{figure}
\centering
\includegraphics[width=1\columnwidth]{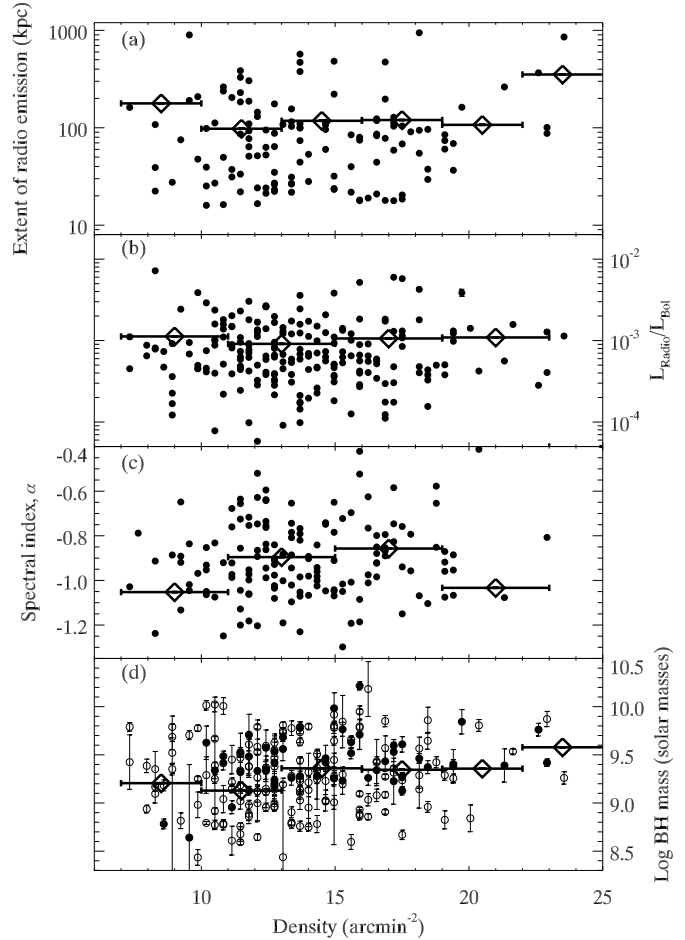}
\caption{Examining relationships between the large-scale environment of the CARLA RLAGN and properties of the radio emission and SMBH. From top to bottom: (a) the extent of the radio emission for resolved CARLA sources; (b) ratio of radio to bolometric AGN power (for the RLQ only); (c) spectral index $\alpha$ (where $S_\nu\propto \nu^{\alpha}$); and (d) SMBH mass for the 211 CARLA AGN matched with the \citet{Shen2011} SDSS catalogue. Solid circles are SMBH masses derived from Mg{\sc ii} whilst open circles are derived from C{\sc iv}. Open diamonds are the mean in each density bin. There is no correlation between RLAGN environment and any radio property, although there is a low-significance (2.3$\sigma$) weak correlation with SMBH mass. \label{fig:radio_prop}}
\end{figure}

\subsection{Observational selection biases}

\begin{table}
\begin{tabular}{llll}
\hline
&Number& Spearman-rank& P \\
&& coefficient&  \\
 \hline
Extent of radio emission &127& 0.03&0.77\\ 
L$_{\rm Radio}$/L$_{\rm Bolometric}$&211&-0.05&0.40\\
Spectral index&158& 0.08&0.29\\
SMBH mass &211&$0.16$ & 0.02\\
\hline
\end{tabular}
\caption{\label{table_rs}Results from Spearman rank correlation test between the listed parameters and the surrounding galaxy density of RLAGN. P is the probability of obtaining the rank coefficient by chance; P values greater than $0.05$ mean the correlation is not significant. The only significant correlation is between the environmental galaxy density and the SMBH mass of the RLAGN. }
\end{table}

In the previous section we showed that RLAGN are generally found in denser environments than radio-quiet galaxies.  
In this section we examine whether there is an observational selection bias that makes it easier for us to observe RLAGN if they are located in dense environments.

\citet{Barthel1996} suggested that radio luminosity is enhanced in rich environments because the relativistic plasma is confined by dense intracluster medium (ICM). This confinement prevents the radio lobes from expanding adiabatically, and a larger fraction of the electron's energy is lost through synchrotron emission rather than expansion losses. Therefore these confined sources are more efficient in transferring AGN power into radio luminosity. In a luminosity-limited sample of radio-loud galaxies, such as the CARLA sample, this effect may create a selection bias. If the relativistic plasma is confined we also expect the source to appear smaller, and the spectral index of their radio emission to steepen due to Fermi acceleration across the shock fronts as the hotspots decelerate \citep{Athreya1998b,Klamer2006}.

We test whether the properties of the radio emission are affected by the ambient gas in rich environments in Fig.\,\ref{fig:radio_prop} by investigating correlations between the environments of the CARLA galaxies and their extent, spectral index and the ratio of their radio to bolometric luminosity (for the quasars). There are two caveats to our method. First, our measure of environment is not robust for individual objects as it is strongly influenced by cosmic variance along the line of sight. Nevertheless, our large sample means this variance should balance out and we are able to measure statistical properties.  Second, complex trends exist between radio extent, spectral index, luminosity  and redshift \citep{Blundell1999}. We try to minimise these issues by using the CARLA sample which is comprised of only the most powerful RLAGN which reside in a narrow 3\,Gyr window ($1.3<z<3.2$), yet these intrinsic relations may make it difficult to find independent correlations with environmental galaxy density.

As shown in Fig.\,\ref{fig:radio_prop}, there is no significant correlation between the radio properties of the RLAGN and its environment. The mean radio extent of resolved sources is $\sim125$\,kpc (median $\sim80$\,kpc) and is approximately constant with density. A Spearman rank correlation test shows that neither the ratio of radio-to-bolometric luminosity, nor the spectral index is correlated with the surrounding galaxy density (see Table\,\ref{table_rs}). There is also no significant difference between the environments of radio sources that are resolved in the FIRST survey and those that are not (KS p=0.16). Taking into account the caveats discussed above, this indicates that the size of the radio emission is not affected by the Mpc-scale environment of the AGN host galaxy. 

There is no evidence that the sizes, spectral indices or radio luminosities of the AGN are affected by their environment, but since all three properties depend on other properties, such as age, orientation and redshift, we are unable to draw strong conclusions from these null results alone. However the literature also suggests that powerful radio emission is not enhanced in dense environments, nor is a dense environment a prerequisite for visible radio emission. \citet{Wylezalek2013} found no correlation between the radio power of the CARLA AGN and their environment over 2 orders of magnitude in radio power, covering $10^{27.5-29.5}$W\,Hz$^{-1}$. Similarly, \citet{Karouzos2014} found no trend between radio power and environmental density of lower-luminosity RLAGN, though at lower redshifts, and \citet{Donoso2010} found a negative trend between radio power and galaxy clustering for radio galaxies. In contrast to these works, \citet{Falder2010} found a trend of increasing galaxy overdensity with increasing AGN radio luminosity for $z\sim1$ RLAGN. These works all look at different ranges of radio power, which may explain some of the differences in the results. Whilst there is not unanimous agreement amongst the literature, the majority of studies suggest that the radio luminosity of a RLAGN is not amplified in dense environments.

Powerful radio jets can also be observed outside dense environments: \citet{Venemans2007} searched for proto-clusters around 8 RLAGN and found that 2 were not in dense environments. Additionally, almost a quarter (24\%)  of CARLA RLAGN reside in average environments: the galaxy densities within 1\arcmin\ of these sources are less than $1\sigma$ above the control field density. Whilst these RLAGN exhibit richer than average environments at radii greater than 1\arcmin\ (Fig.\,\ref{fig:radial_profile}), the radio emission from all CARLA RLAGN is limited to within 1\arcmin\ of the host galaxy and therefore is unlikely to be greatly affected by the gas associated with these more distant structures. This implies that dense ambient gas is not a prerequisite for high-redshift radio-jets to be detected.

We conclude that radio emission from the CARLA sample is not significantly or systematically amplified when a CARLA source lies within a dense environment, and therefore there is no observational selection bias that makes it easier for us to observe RLAGN when they are located in richer environments. Since we cannot invoke an observational selection bias to explain the results from Section \ref{section3.1} (that RLAGN reside in denser environments than radio-quiet galaxies), we resort to the alternative explanation: the probability of the central SMBH launching a radio-emitting jet depends on the host galaxy's environment.

If the large-scale environment influences the properties of the SMBH directly we may expect a correlation between the SMBH mass and its environment.   Although there is a great deal of scatter in Fig.\,\ref{fig:radio_prop}d, we find a mildly significant, but small tendency for more massive SMBHs to lie in denser environments: the Spearman correlation coefficient is $0.16$ at 98\% confidence. This correlation may simply result from the relation between stellar and SMBH mass  \citep{Ferrarese2000} and the tendency for high mass galaxies to reside in denser environments than low mass galaxies. Therefore further investigation of this trend requires a study of the  environment and SMBH mass of RLAGN at fixed stellar mass. Because the significance of the correlation is so low, we do not discuss the result further, however even if this correlation is real, it cannot explain why RLAGN tend to reside in dense environments. Powerful radio emission from AGN require the SMBHs to have masses greater than $10^8$\Msun, but there is no strong correlation between black hole mass and radio luminosity \citep{McLure2004}.   

\section{Discussion}
\label{discussion}
\subsection{Why do RLAGN tend to reside in proto-clusters?}
The main goal of this study is to examine why RLAGN are commonly found in proto-clusters. We showed that RLAGN reside in denser environments than similarly massive radio-quiet galaxies, which suggests a connection between the Mpc-scale environment and radio-loudness. We also showed that this connection is not due to the ICM environment amplifying the radio emission. Our results therefore suggest that the {\em launching} of powerful radio jets depends in some way on the environment of the host galaxy. We can, however, only speculate on ways the Mpc-scale environment can influence the properties of the AGN on sub-pc scales.

The semi-analytic galaxy formation models of \citet{Fanidakis2011} suggest that galaxy mergers are very important for determining whether an AGN becomes radio-loud. \citet{Fanidakis2011} were able to reproduce the radio luminosity function as long as the black holes accreted gas in a chaotic manner \citep[which keeps the spin of most black holes low;][]{King2006}, but the spin of the most massive black holes was driven by black hole mergers (which themselves are driven by the galaxy merger rate). These black hole mergers result in rapidly spinning black holes, which may be more likely to launch powerful jets \citep[e.g.,][]{Blandford1977}. We expect high galaxy merger rates in proto-clusters because of the high galaxy densities and the relatively low velocity dispersions compared to virialized clusters \citep{Venemans2007,Shimakawa2014a}. Indeed, \citet{Lotz2011} found the merger rate in a $z=1.62$ proto-cluster was $\sim$10 times higher than the coeval field rate. High galaxy merger rates imply high black hole merger rates, and therefore high black hole spins, which may explain why proto-clusters commonly host RLAGN.

However, the radio-loudness of an AGN depends on more than just the spin of the black hole; the accretion mode of the AGN is also important \citep{Tchekhovskoy2010}. \citet{Fernandes2011} found a correlation between jet power and accretion rate, which implies that high accretion rates are required to generate the  powerful jets of the CARLA RLAGN. Frequent galaxy mergers in the proto-cluster environment will dump cold gas into the RLAGN host galaxy. Some fraction of this gas is likely to have sufficiently low angular momentum to fall into the nucleus, leading to high accretion rates on to the black hole. Thus the proto-cluster environment may promote high accretion rates on sub-pc scales, which are required to generate powerful RLAGN.

\subsection{Fraction of massive galaxies that become powerful RLAGN}
The environments of radio-quiet massive galaxies differ from those of RLAGN, which means only a subset of massive galaxies undergo radio-loud feedback as powerful as  L$_{\rm 500MHz}=10^{27.5}$\,W\,Hz$^{-1}$. We estimate this fraction by scaling the CARLA probability density histogram in Fig.\,\ref{main}, using a least squares fitting algorithm, so it matches the height of the control histogram at high densities (14-20 galaxies per arcmin$^{-2}$). We find that 50\% of the control galaxies have similar environments to the CARLA RLAGN\footnote{We obtain similar results when we run the algorithm using different density bin sizes, and when we use the full CARLA sample or the CARLA radio galaxy subsample.}. Therefore only $50$\% of the massive control galaxies can go through a  RLAGN phase as powerful as L$_{\rm 500MHz}=10^{27.5}$\,W\,Hz$^{-1}$ during $1.3<z<3.2$. 

A major limitation of this result is that we only discuss the most powerful radio-loud sources at $1.3<z<3.2$, rather than all RLAGN. We cannot constrain the fraction of massive galaxies that have RLAGN feedback, only the fraction that undergo the most powerful feedback usually associated with FRII radio galaxies. 

\subsection{Total lifetime of the radio-emitting phase}
The proportion of time a RLAGN host galaxy is observed as radio-loud can be estimated through the ratio of RLAGN to potential host galaxies: 
\begin{equation}
f=\frac{n(RLAGN)}{n({\rm host}) \times f_{RL}}
\end{equation}
where $f_{RL}$ is the fraction of galaxies that can become radio-loud ($<0.5$), $n$(host) is the  number density of galaxies that have similar masses as the RLAGN galaxies, and $n(RLAGN)$ is the number density of RLAGN with L$_{\rm 500MHz}>10^{27.5}$\,W\,Hz$^{-1}$, which is $5-10\times10^{-8}$\,Mpc$^{-3}$ comoving  at $1.3<z<3.2$  \citep{Rigby2011}. 

A large uncertainty in this calculation comes from $n$(host). The stellar mass function is so steep at the high mass end that $n$(host) is completely dependent on the number density of the most massive hosts (i.e., galaxies with $\gtsim10^{11.5}$\,\Msun). The number density of $>10^{11.5}$\Msun\ galaxies is $\sim1-7\times10^{-6}$\,Mpc$^{-3}$  \citep{Santini2012,Muzzin2013b}, and does not appear to evolve rapidly across the $1.3<z<3.2$ redshift interval. However, the uncertainties on this number density are large because cosmic variance has a strong impact on measuring the density of such rare sources.

Using equation 1 we estimate that a powerful RLAGN is radio-bright for at least 2 per cent of the time between $z=3.2$ and $z=1.3$, i.e.~$\gtsim60$\,Myr. The radio-bright lifetime of RLAGN is limited to a few$\times10$\,Myr \citep{Blundell1999} so this result is consistent with galaxies undergoing a single powerful RLAGN during this period. However,   since we derive lower limits of the total radio-emitting lifetime we cannot rule out that powerful radio-loud AGN feedback is a reoccurring phenomenon with each RLAGN host galaxy undergoing multiple powerful radio-loud episodes between $z=3.2$ and $z=1.3$. 

\subsection{Heating the intracluster medium}

Approximately half of CARLA RLAGN reside in rich environments containing IRAC-selected overdensities of $>2\sigma$ significance \citep{WylezalekLF}. These environments are highly likely to be proto-clusters or clusters, several of which have already been confirmed \citep{Venemans2007,Galametz2010a,Hatch2011b,Galametz2013}. However, Fig.\,\ref{fig:radial_profile} shows that even the RLAGN which have low densities in the central arcmin have richer environments at larger radii ($>0.5$\,Mpc). Therefore it is possible that the majority of RLAGN reside in clusters and proto-clusters.

The maximum number density of galaxies that become radio-loud is a few$\times10^{-6}$\,Mpc$^{-3}$ and is therefore similar to that of $>10^{14}$\Msun\ galaxy clusters in the present day \citep{Vikhlinin2009}. It is therefore possible that every galaxy cluster hosted a powerful RLAGN at some point between $1.3<z<3.2$. 

The minimum radio-loud lifetime allows us to estimate the total energy deposited into the forming ICM. The energy output by powerful RLAGN at $z\ge2$ is a few~$\times10^{59}$\,erg in $30$\,Myr considering only the electron population, and up to $10^{62}$\,erg when including the protons in the jets  \citep{Erlund2006,Johnson2007,Erlund2008}. 

It is well known that extra heating of the ICM on top of gravitational heating, of about 0.5-1\,keV per particle, is required to reproduce the observed excess of ICM entropy  \citep{Kravtsov2012}.  In a cluster of gas mass $1\times10^{14}$\Msun\ this would require $1-2\times10^{62}$\,erg. So the energy deposited by the RLAGN across their active lifetime is sufficient to provide a large fraction of this energy to the ICM. If jets are able to couple efficiently with the surrounding gas, then these powerful early feedback events can have a dramatic and long-lasting impact on the intracluster gas. Heating the intracluster gas may choke the proto-cluster galaxies by cutting off the gas supply and limiting further stellar and SMBH growth \citep{Rawlings2004}.

\section{Conclusions}

\label{conclusions}
We have shown that the environments of high-redshift RLAGN and similarly massive radio-quiet galaxies differ so significantly that less than half of the massive galaxy population (with masses in the range $10^{11-11.5}$\,M$_{\odot}$) at $1.3<z<3.2$ may host a  powerful RLAGN.  Even when we compare the subset of RLAGN and control galaxies that have the same density within $\sim0.5$\,Mpc, we find the RLAGN have richer environments beyond this radius, implying that they reside in larger structures. Thus having a high stellar mass does not ensure that a galaxy will host a RLAGN and we suggest that the presence of a radio-loud jet may be influenced by the Mpc-scale environment of the host galaxy.

The correlation between radio-loud AGN activity and high density environments is not the result of an observational selection bias. We examined whether dense environments amplifies the radio emission from a RLAGN by confining the radio emitting plasma or transforming AGN power into radio power. We found no correlation between the environment and the radio size, spectral index or fraction of AGN power emitted at radio frequencies.  In addition, approximately a quarter of RLAGN appear to reside in average environments on 0.5\,Mpc scales (which is typically the size of the radio emission). We therefore conclude that dense environments do not greatly amplify the radio emission from a RLAGN at this redshift, but the probability that a jet is launched from the host galaxy is likely to depend on the Mpc-scale environment.

We estimate the maximum space density of galaxies that experience a radio-loud episode in the epoch at $1.3<z<3.2$ to be a few$\times10^{-6}$\,Mpc$^{-3}$, which is similar to the space density of proto-clusters: objects that can collapse to form a $>10^{14}$\Msun\ galaxy cluster by the present day. Distant clusters and proto-clusters are reliably found near powerful radio-loud AGN \citep{Venemans2007, Galametz2010a,Galametz2013,Hatch2011a,Hatch2011b,Wylezalek2013} so it is possible that every cluster progenitor experienced a powerful feedback episode during $1.3<z<3.2$. We estimate that each powerful radio-loud galaxy is active for at least $60$\,Myr, and if the jets contain protons, then this feedback could provide enough energy per gas particle to pre-heat the forming ICM.

\section{Acknowledgments}
We sincerely thank the referee for providing useful and constructive comments which improved this paper. NAH acknowledges support from STFC through an Ernest Rutherford Fellowship. NS is the recipient of an ARC Future Fellowship. MJJ thanks the South African SKA for support. This work is based on observations made with the {\it Spitzer Space Telescope}, which is operated by the Jet Propulsion Laboratory, California Institute of Technology.
\bibliographystyle{mn2e}\bibliography{HzRG-bib,mn-jour}
\label{lastpage}
\clearpage
\end{document}